\documentclass[12pt]{article}
\usepackage[utf8]{inputenc}
\usepackage[english]{babel}
\usepackage{amssymb}
\usepackage{amsmath}
\usepackage{graphicx}
\usepackage[small,sc,center]{titlesec}
\usepackage{indentfirst}
\usepackage{siunitx}

\newcommand{\nc}{\newcommand}
\nc{\mzams}{M_\mathrm{ZAMS}}
\nc{\tev}{t_\mathrm{ev}}

\begin{document}

\begin{center}
\textbf{Nonlinear pulsations of stars with initial mass $3\boldsymbol M_\odot$\\
        on the asymptotic giant branch}

\vskip 3mm
\textbf{Yu. A. Fadeyev\footnote{E--mail: fadeyev@inasan.ru}}

\textit{Institute of Astronomy, Russian Academy of Sciences, Pyatnitskaya ul. 48, Moscow, 119017 Russia} \\

Received March 29, 2016
\end{center}

\textbf{Abstract} ---
Pulsation period changes in Mira type variables are investigated using
the stellar evolution and nonlinear stellar pulsation calculations.
We considered the evolutionary sequence of stellar models with
initial mass $\mzams = 3M_\odot$ and population~I composition.
Pulsations of stars in the early stage of the asymptotic giant branch
are shown to be due to instability of the fundamental mode.
In the later stage of evolution when the helium shell source becomes
thermally unstable the stellar oscillations occur in either the fundamental
mode (for the stellar luminosuty $L < 5.4\times 10^3L_\odot$)
or the first overtone ($L > 7\times 10^3L_\odot$).
Excitation of pulsations is due to the $\kappa$--mechanism in the
hydrogen ionization zone.
Stars with intermediate luminosities
$5.4\times 10^3L_\odot < L < 7\times 10^3L_\odot$
were found to be stable against radial oscillations.
The pulsation period was determined as a function of evolutionary time
and period change rates $\dot\Pi$ were evaluated for the first ten
helium flashes.
The period change rate becomes the largest in absolute value
($\dot\Pi/\Pi\approx -10^{-2}~\text{yr}^{-1}$)
between the helium flash and the maximum of the stellar luminosity.
Period changes with rate $|\dot\Pi/\Pi|\gtrsim 10^{-3}~\text{yr}^{-1}$
take place during $\approx 500$~yr, that is nearly one hundredth of
the interval between helium flashes.


\newpage
\section*{introduction}

$o$~Cet type variables (or Miras) belong to numerous pulsating variables
with visual amplitudes greater than two magnitudes and oscillation periods
$10^2~\text{day}\lesssim\Pi\lesssim 10^3~\text{day}$.
The long--term observations spanning one or several decades reveal that
pulsation periods of some Miras undergo secular variation
(either decrease or increase).
One of the first reports on detection  of secular period variation in
the variables $o$~Cet and $\chi$~Cyg was made by Eddington and Plakidis (1929).
Recent estimates of the fraction of Miras with unambiguously detected
period changes are of a few percent of their total number
(Templeton et al., 2005; Lebzelter and Andronache, 2011),
whereas the rate of period change is as high as a few tenths
of the percent per year.
For example, the pulsation period of T~UMi ($\Pi=285$~day) decreases with
rate of $d\ln\Pi/dt=-8.47\times 10^{-3}~\text{yr}^{-1}$
and the period of LX~Cyg ($\Pi=520$~day) increases with rate of
$d\ln\Pi/dt=6.47\times 10^{-3}~\text{yr}^{-1}$
(Templeton et al., 2005).
It should be noted that detection of secular period change in Miras
is hindered by small (i.e. with amplitude as high as 10\%)
random period changes of unknown nature
(Percy and Colivas, 1999; Koen and Lombard, 2004; Uttenthaler et al., 2011).

Miras are assumed to be the stars of the asymptotic giant branch (AGB)
with nuclear energy sources in the hydrogen--burning and helium--burning
shells surrounding a carbon--oxygen core
(Wood and Zarro, 1981; Iben and Renzini, 1983; Boothroyd and Sackmann, 1988;
Vassiliadis and Wood, 1993).
Shell helium burning is characterized by thermal instability, that is by
short--term enhancements of nuclear energy release
(Schwarzschild and H\"arm, 1965; Weigert, 1966).
The thermal flash amplitude is so large that the energy release
in the helium burning shell is by several orders of magnitude higher
than the surface luminosity.
The short--lived intershell convection zone created due to the large
energy flux transports the products of helium burning upwards.
After a shorth time the innner boundary of the outer convection zone
descends in these layers and dredges up the material enriched in
heavier elements to the surface of the red supergiant.
Detection of  unstable technetium isotopes ${}^{97}$Tc and ${}^{98}$Tc
(Merrill, 1952;
Little--Marenin and Little, 1979;
Lebzelter and Hron, 1999)
with half--life time $\approx 4.2\times 10^6$~yr
corroborates that Miras are thermally pulsing AGB stars.

Helium flashes are accompanied by changes in stellar radius and luminosity
on the time scale of $\sim 10^3$~yr and are thought to be responsible for secular
period variation in Miras.
In particular, the duration of the surface luminosity changes due to
the helium flash is from one to a few hundredths of the time interval between
two thermal pulses and is comparable with fraction of Miras with
secular period variation.
Moreover, Wood and Zarro (1981) showed that the period change rates
in Mira type variables R~Hya, R~Aql and W~Dra agree with lumiunosity
variations predicted by evolutionary calculations.
This conclusion, however, implies that these stars are the first
overtone pulsators and, as noted the authors of the work, should be
corroborated by the methods of the stellar pulsation theory.

The pulsation mode of Miras remains disputable.
For example, Wood and Sebo (1996), Weiner (2004), Perrin et al. (2004)
argued in favor of the fundamental mode, whereas
Tuchman (1991), Tuthill et al. (1994), Haniff et al. (1995)
showed that Miras are the first overtone pulsators.
It should be noted that these conclusions are based
on measurements of angular diameters of a few stars with known
parallaxes.
Results of photometric observations of many long period variables
in the Large and Small Magellanic Clouds (Ita et al. 2004) and in the
dwarf galaxies NGC~147 and NGC~185 
(Lorenz et al., 2011) allow us to conclude that Miras pulsate in the both
fundamental mode and first overtone.

Below we discuss the results of hydrodynamic computations of radial pulsations
of population~I stars with initial mass $\mzams=3M_\odot$ on the early AGB
stage as well as on the advanced stage of thermal instability of
the helium shell source.
The goal of the study is to investigate the evolutionary changes of
pulsational properties of Miras.
In particular, we specially pay attention to the mode of radial oscillations
and to changes in the pulsation period.
Comparison of theoretical period change rates with observational data
is of importance to corroborate the evolutionary status of Miras
as AGB stars.

\section*{calculations of stellar evolution}

To solve the equations of radiation  hydrodynamics for self--excited
stellar pulsations we need to impose the initial conditions corresponding
to the hydrostatic equilibrium.
In the present study the initial conditions were represented as models
of the evolutionary sequence calculated with the MESA code version 7624
(Paxton et al., 2011; 2013; 2015).
The initial chemical composition was assumed to be that of population~I stars
with fractional mass abundances of hydrogen and elements heavier than helium
$X=0.7$ and $Z=0.02$, respectively.
The reaction network included isotopes from ${}^{1}\mathrm{H}$ to
${}^{26}\mathrm{Mg}$.

Convection was treated through the classical mixing length theory
(B\"ohm--Vitense, 1958) with a mixing length to pressure scale height ratio
$\alpha_{\mathrm{MLT}}=1.8$.
Convective overshooting was taken into account using the prescription
of Herwig (2000) and the coefficient of convective diffusion $D_\mathrm{conv}$
was assumed to reduce exponentially as a function of geometric distance $z$
from the edge of the convective zone:
\begin{equation}
\label{dconv}
D_\mathrm{conv} = D_0 \exp\left(-\frac{2z}{f H_\mathrm{P}}\right) .
\end{equation}
Here $H_\mathrm{P}$ is the local pressure scale height,
$D_0$ is the coefficient of convective diffusion within the
convective zone in the layer located at $4\times 10^{-3} H_\mathrm{P}$
from the boundary of convective stability.
Relation (\ref{dconv}) was used for smoothing of the convective diffusion
coefficient at boundaries of all convective zones and the dimensionless
parameter was $f=0.016$.

The mass loss rate $\dot M$ in units of $M_\odot/\text{yr}$
was evaluated according to Bl\"ocker (1995):
\begin{equation}
 \dot M = 4.83 \times 10^{-9} \eta_\mathrm{B} (M/M_\odot)^{-2.1} (L/L_\odot)^{2.7}
          \dot M_\mathrm{R} ,
\end{equation}
where
\begin{equation}
\dot M_\mathrm{R} = 4\times 10^{-13} \eta_\mathrm{R} (L/L_\odot) (R/R_\odot) (M/M_\odot)^{-1}
\end{equation}
is the Reimers (1975) formula.
The adopted free parameters are
$\eta_\mathrm{B}=0.1$ and $\eta_\mathrm{R}=1$.

\begin{figure}
\centerline{\includegraphics[width=15cm]{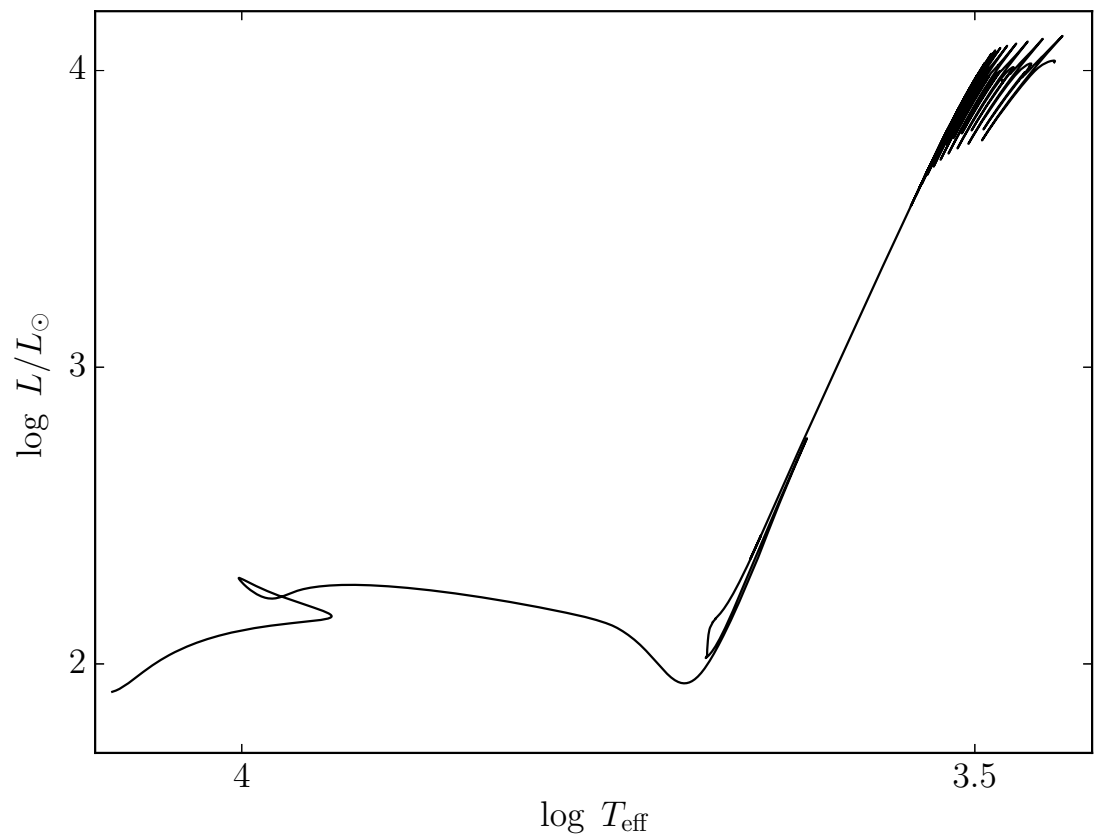}}
\caption{The evolutionary track of the star $\mzams=3M_\odot$ from the
         zero age main sequence to the thermally pulsing AGB.}
\label{fig1}
\end{figure}

Fig.~\ref{fig1} displays the evolutionary track of the star with initial mass
$\mzams=3M_\odot$ on the Hertzsprung--Russel diagram (HRD).
The initial point of the track corresponds to the homogeneous
chemical composition along the radius.
At the final point of the track the age, mass of the star and mass
of the carbon--oxygen core are $4.75\times 10^8$~yr, $M=2.25M_\odot$ and
$M_\mathrm{CO}=0.61M_\odot$, respectively.
Abrupt changes of the luminosity in the upper part of the track
within $3.5 < \log (L/L_\odot) < 4.1$ are due to thermal instability
of helium shell burning.
The evolutionary computations of the present study comprise the first ten
helium flashes.

\begin{figure}
\begin{picture}(300,300)
\put(50,150){\includegraphics[width=15cm]{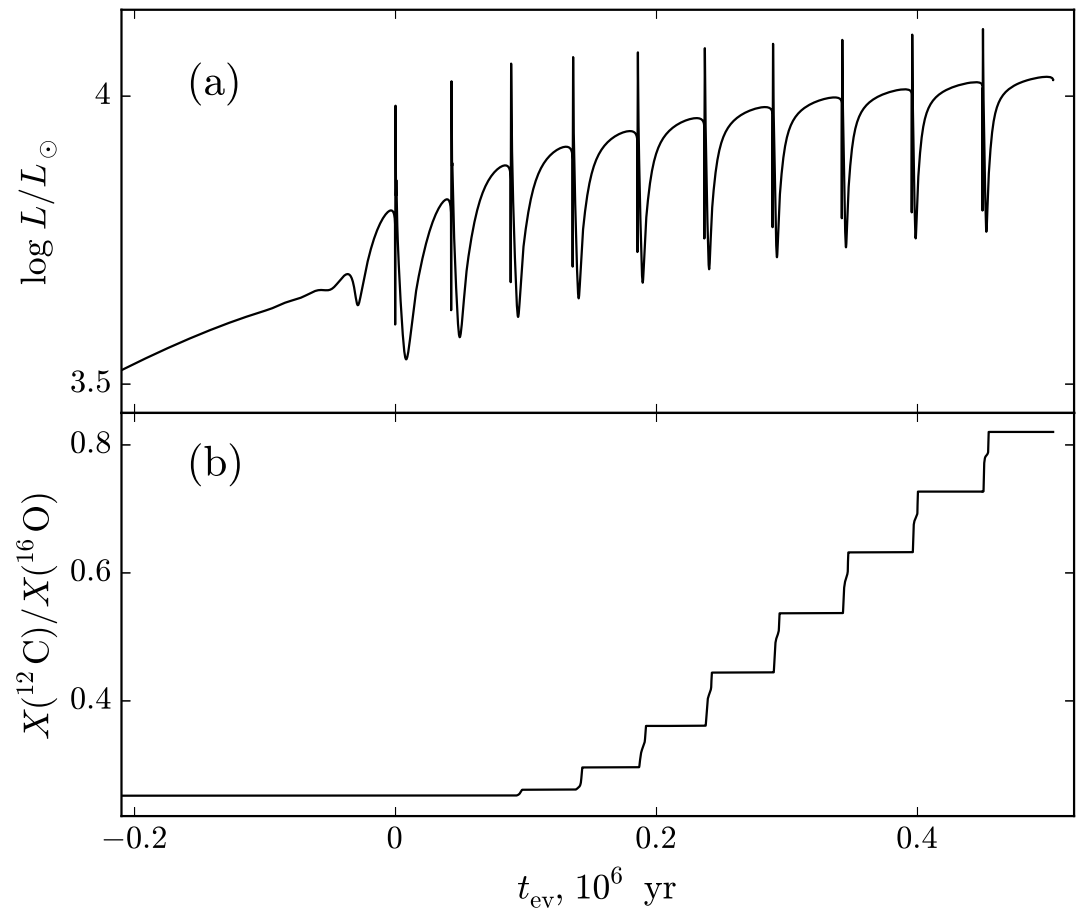}}
\end{picture}
\caption{Temporal dependences of the stellar luminosity (a) and the ratio
         of surface mass abundances of carbon and oxygen (b) during
         the AGB stage.}
\label{fig2}
\end{figure}

The temporal dependence of the stellar luminosity $L$ is shown in Fig.~\ref{fig2}a
where for the sake of convenience the evolutionary time $\tev$ is set to zero
at the first helium flash luminosity maximum.
As the star evolves the time interval between helium flashes gradually
increases from $4.3\times 10^4$~yr to $5.3\times 10^4$~yr.
Results of our evolutionary computations agree with those by Herwig (2000)
notwithstanding the fact that during the last decade and a half the data
on equation of state, opacity and thermonuclear reaction rates
underwent substantial changes.

Thermal flashes in the helium burning shell and convective dredge up lead to
enrichment of outer stellar layers in helium, carbon and oxygen.
This is illustrated in Fig.~\ref{fig2}b where the carbon to oxygen
surface abundance ratio $X({}^{12}\mathrm{C})/X({}^{16}\mathrm{O})$
is shown as a function of $\tev$.

The initial conditions for hydrodynamic computations of nonlinear stellar
pulsations were obtained from interpolation of the radius $r(M_r)$,
luminosity $L_r(M_r)$, total pressure $P(M_r)$ and temperature $T(M_r)$
represented as a function of mass coordinate
$M_r = 4\pi\int_0^r x^2 \rho(x) dx$, where $\rho$ is the gas density.
To diminish interpolation errors the evolutionary computations were
done with the number of mass zones $\sim 5\times 10^4$, so that
the interpolation region spanned $\sim 10^4$ outermost zones of the
evolutinary model.
The inner boundary of the hydrodynamical model was set in the layers with
radius
$10^{-3}\lesssim r/R\lesssim 10^{-2}$, where $R$ is the radius of the
outer boundary of the evolutionary model.
All hydrodynamical models were computed with the number of
Lagrangian mass zones $N=10^3$.
The mass intervals increase geometrically inwards.
The inner boundary of the hydrodynamical model locates near the outer boundary
of the carbon--oxygen core where the temperature and pressure gradients
abruptly grow inwards.
To diminish approximation errors of the difference equations
in this region of the hydrodynamical model we used the mesh with 100 innermost
equally spaced mass intervals.

The envelope of the pulsating star was assumed to be chemically homogeneous
and calculation of thermodynamic quantities in numerical solution
of the equations of hydrodynamics was carried out as bicubic interpolation of
tabular data with respect to $\log T$ and $\log R = 18 - 3\log T + \log\rho$.
The tables of thermodynamic quantities were calculated for each hydrodynamical
model for abundances of the outer layers of the evolutionary model.
The equation of state tables were calculated with the library of programs
FreeEOS (Irwin, 2012).
Thermodynamic quantities calculated in such a way were found to agree
with those obtained from MESA computations.

\section*{hydrodynamical models}

In the present study the equations of radiation hydrodynamics for nonlinear
stellar pulsations are solved together with transport equations for
time--dependent convection in spherically--symmetric geometry
(Kuhfu\ss, 1986).
The system of differential equations and the adopted parameters of the
convection model are given in our previous paper(Fadeyev, 2015).
The solution of the equations of hydrodynamics is represented by
physical quantities as a function of mass coordinate $M_r$
and time $t$.
Time $t=0$ corresponds to the initial state of hydrostatic equilibrium.

As in the studies of nonlinear pulsations of massive red supergiants
(Fadeyev, 2012; 2013) the mean pulsation period $\Pi$ was evaluated by
a discrete Fourier transform of the kinetic energy $E_\mathrm{K}(t)$
of pulsational motions of the stellar enevelope on the time interval
spanning several hundred pulsation cycles.
The instability growth rate $\eta = \Pi d\ln E_\mathrm{K,max}/dt$
was determined from maximum values of the kinetic energy on the initial
integration interval where $\ln E_\mathrm{K,max}$ changes linearly with
time $t$.
An inverse of the growth rate is the number of pulsation cycles
corresponding to either increase or decrease of the maximum
kinetic energy by a factor of $e=2.718\ldots$.

Fig.~\ref{fig3} shows the plot of evolutionary change of the stellar
luminosity $L$ in vicinity of the first helium flash.
On the same figure we give the mean values of the luminosity of
hydrodynamical models.
For pulsationally unstable ($\eta > 0$) and pulsationally stable ($\eta < 0$)
models they are shown by filled and open circles, respectively.

The most striking feature of the plots displayed in Fig.~\ref{fig3} is
stability against radial pulsations of stars with luminosity
$6000L_\odot < L < 8700L_\odot$ and pulsational instability at luminosities
beyond this interval.
The discrete Fourier transform of radii $r_j$
of the hydrodynamical models allows us to conclude that stars
with luminosity $L < 6000L_\odot$ pulsate in the fundamental mode
whereas stars with $L > 8700L_\odot$ are the first overtone oscillators.
The difference in pulsation modes is illustrated in Fig.~\ref{fig4} where
the radial dependences of the normalized power spectral density of the
radial displacement $|\Delta r/\Delta R|$ are shown for two models indicated
in Fig.~\ref{fig3} as f (fundamental mode) and h1 (first overtone).
The independent variable of the plot in Fig.~\ref{fig4} is the ratio
of the mean radius
$\bar{r}_j$ of the $j$--th mass zone to the mean radius of the outer
boundary $\bar{r}_N$ of the hydrodynamical model.

Pulsations in the fundamental mode arise during the early AGB stage
as well as on the stage of the thermal instability of the helium shell
source.
The upper luminosity boundary for pulsations in the fundamental mode
decreases with evolutionary time $\tev$ from $L\approx 6\times 10^3L_\odot$
near the first helium flash to $L\approx 5\times 10^3L_\odot$
near the sixth helium flash.
After the seveth helium flash the stellar luminosity exceeds the upper
luminosity limit for the fundamental mode and further stellar evolution
is accompanied by oscillations in the first overtone or by pulsational stability.

In the hydrodynamical model indicated in Fig.~\ref{fig3} as h1
the mean values of the radius and temperature at the node of the first overtone
are $\bar{r}_j/\bar{r}_N=0.76$ and $\bar{T}\approx 1.6\times 10^4$~K, respectively.
The region of excitation of pulsational instability in the first overtone
(i.e. the layers with the mechanical work per closed cycle $\oint PdV > 0$,
where $V=1/\rho$ is the specific volume) locates in layers with radii
$\bar{r}\ge 0.84\bar{R}$ and temperatures $\bar{T}\le 1.2\times 10^4$~K.
Therefore, the first overtone oscillations are driven in the hydrogen
ionization zone.
Fundamental mode pulsations seem to be driven in the both hydrogen and
first helium ionization zone since the condition $\oint PdV > 0$
is fulfilled in layers with temperature $\bar{T} \le 1.7\times 10^4$~K.

The radius of the model h1 pulsating in the first overtone with period of
$\Pi=176$~day is almost twice the radius of the model f pulsating in
the fundamental mode ($\Pi=96$~dat).
Nevertheless, the temporal dependences of the bolometric light
$M_\mathrm{bol}$ and the velocity of the outer boundary $U$ do not show
significant differences in the oscillation amplitude and the shape
of the curve (Fig.~\ref{fig5}).
At the same time the hydrodynamical models of fundamental mode and
first overtone pulsators reveal significant differences on the period--luminosity
and period--radius diagrams (Fig.~\ref{fig6}).
Periods of fundamental mode pulsations are $\Pi < 158$~day and
for first-overtone pulsations $\Pi > 144$~day.

For the fundamental mode pulsators the period--luminosity and
period--radius relations can be approximated as
\begin{gather}
\label{p-l_f}
 \log L/L_\odot = 2.42 + 0.595 \log\Pi ,
\\
\label{p-r_f}
 \log R/R_\odot = 1.40 + 0.428 \log\Pi ,
\end{gather}
and for first--overtone pulsators
\begin{gather}
\label{p-l_h1}
\log L/L_\odot = 2.04 + 0.855 \log\Pi ,
\\
\label{p-r_h1}
\log R/R_\odot = 0.77 + 0.781 \log\Pi ,
\end{gather}
where the pulsation period $\Pi$ is expressed in days.

\section*{pulsation period change}

Fig.~\ref{fig7}a shows the temporal dependence of the stellar luminosity $L$
for the evolutionary time interval spanning three (from seventh to ninth)
helium flashes.
As in Fig.~\ref{fig3} the mean values of the luminosity of hydrodynamical
models are shown for both positive and negative growth rates.
In contrast to the first helium flash all unstable models are the
first overtone pulsators.

Fig.~\ref{fig7}b shows for the same evolutionary time interval
the plot of the pulsation period $\Pi$ with gaps for $\eta < 0$.
Time intervals between thermal flashes are
$5.30\times 10^4$ and $5.36\times 10^4$~yr
but the most significant changes of the period occur within
shorter time interval ($\lesssim 10^4$ yr)
in vicinity of the maximum light.

The rate of period change was calculated using the numerical didderentiation.
To obtain typical estimates of $\dot\Pi$ we choosed the eighth helium flash
and calculated nearly thirty hydrodynamical models.
Results of these computations are illustarted in Fig.~\ref{fig8} where
temporal dependences of
the pulsation period $\Pi$ and the rate of period change $\dot\Pi/\Pi$
are shown for the evolutionary time interval $\approx 1.3\times 10^3$~yr.
It is seen that the rate of period change is greater in absolute value
than 0.1\% within the evolutionary time interval $\approx 500$~yr,
that is during one hundredth of the interval between helium flashes.
The period change rate becomes the largest in absolute value
($|\dot\Pi/\Pi|\approx 10^{-2}$)
during the abrupt stellar radius decrease before the maximum of
stellar luminosity but so great secular period variation continues no longer
than a hundred years.

\section*{conclusion}

The results of computations presented above allow us to conclude that
radial pulsations of long period variables with masses
$2.2M_\odot < M < 3M_\odot$
arise in both the fundamental mode and the first overtone.
An unexpected conclusion is that the models demonstrated existence of the
region of pulsational stability separating stellar oscillations in
different modes.
Therefore, together with secular changes of the pulsation period
one might observe secular changes of the amplitude of light curves.
The change of the pulsation amplitude should be a more rare phenomenon
since it occurs on a shorter time scale.
Nevertheless, secular amplitude decrease observed in Mira type variables R~Cen
(Hawkins et al., 2001) and T~UMi (Szatm\'ary et al., 2003) seems to be
such a case.

Theoretically calculated period change rates obtained in the present study
agree with those observed in Miras.
However one shoul bear in mind that results of our computations
comprise the small part of AGB stars.
Therefore it is of importance to continue the consistent evolutionary
and stellar pulsation computations for Miras with various initial
masses $\mzams$ for longer evolutionary tracks.

\section*{references}

\begin{enumerate}

\item T. Bl\"ocker, Astron. Astrophys. \textbf{297}, 727 (1995).

\item E. B\"ohm--Vitense, Zeitschrift f\"ur Astrophys. \textbf{46}, 108 (1958).

\item A.I. Boothroyd and I.--J. Sackmann, Astrophys. J. \textbf{328}, 632 (1988).

\item A.S. Eddington and S. Plakidis, MNRAS \textbf{90}, 65 (1929).

\item Yu.A. Fadeyev, Pis'ma Astron. Zh. \textbf{38}, 295 (2012)
      [Astron. Lett. \textbf{38}, 260 (2012)].

\item Yu.A. Fadeyev, Pis'ma Astron. Zh. \textbf{39}, 342 (2013)
      [Astron.Lett. \textbf{39}, 306 (2013)].

\item Yu.A. Fadeyev, MNRAS \textbf{449}, 1011 (2015).

\item C.A. Haniff, M. Scholz, and P.G. Tuthill, MNRAS \textbf{276}, 640 (1995).

\item G. Hawkins, J.A. Mattei, and G. Foster, PASP \textbf{113}, 501 (2001).

\item F. Herwig, Astron. Astrophys. \text{360}, 952 (2000).

\item I. Iben and A. Renzini, Annual Rev. Astron. Astrophys. \textbf{21}, 271 (1983).

\item A.W. Irwin, Astrophysics Source Code Library, record ascl:1211.002 (2012).

\item Y. Ita, T. Tanab\'e, N. Matsunaga, et al., MNRAS \textbf{347}, 720 (2004).

\item C. Koen and F. Lombard, MNRAS \textbf{353}, 98 (2004).

\item R. Kuhfu\ss, Astron. Astrophys. \textbf{160}, 116 (1986).

\item T. Lebzelter and S. Andronache, IBVS \textbf{5981}, 1 (2011).

\item T. Lebzelter and J. Hron, Astron. Astrophys. \textbf{351}, 533 (1999).

\item I.R. Little--Marenin and S.J. Little, Astron.J. \textbf{84}, 1374 (1979).

\item D. Lorenz, T. Lebzelter, W. Nowotny, et al., Astron. Astrophys. \textbf{532}, A78 (2011).

\item P.W. Merrill, Astrophys. J. \textbf{116}, 21 (1952).

\item B. Paxton, L. Bildsten, A. Dotter, et al., Astropys. J. Suppl. Ser. \textbf{192}, 3 (2011).

\item B. Paxton, M. Cantiello,  P. Arras, et al., Astropys. J. Suppl. Ser. \textbf{208}, 4 (2013).

\item B. Paxton, P. Marchant, J. Schwab, et al., Astropys. J. Suppl. Ser. \textbf{220}, 15 (2015).

\item J.R. Percy and T. Colivas, Publ. Astron. Soc. Pacific \textbf{111}, 94 (1999).

\item G. Perrin, S.T. Ridgway, B. Mennesson, et al., Astron. Astrophys. \textbf{426}, 279 (2004).

\item D. Reimers, \textit{Problems in stellar atmospheres and envelopes}
      (Ed. B. Baschek, W.H. Kegel, and G. Traving, New York: Springer-Verlag, 1975), p. 229.

\item M. Schwarzschild and R. H\"arm, Astrophys. J. \textbf{142}, 855 (1965).

\item K. Szatm\'ary, L.L. Kiss, and Zs. Bebesi, Astron. Astrophys. \textbf{398}, 277 (2003). 

\item M.R. Templeton, J.A. Mattei, and L.A. Willson, Astron. J. \textbf{130}, 776 (2005).

\item Y. Tuchman, Astrophys. J. \textbf{383}, 779 (1991).

\item P.G. Tuthill, C.A. Haniff, J.E. Baldwin, and M.W. Feast, MNRAS \textbf{266}, 745 (1994).

\item S. Uttenthaler, K. van Stiphout, K. Voet, et al., Astron. Astrophys. \textbf{531}, A88 (2011).

\item E. Vassiliadis and P.R. Wood, Astrophys. J. \textbf{413}, 641 (1993).

\item A. Weigert, Zeitschrift f\"ur Astrophys. \textbf{64}, 395 (1966).

\item J. Weiner, Astrophys. J. \textbf{611}, L37 (2004).

\item P.R. Wood and K.M. Sebo, MNRAS \textbf{282}, 958 (1996).

\item P.R. Wood and D.M. Zarro, Astrophys. J. \textbf{247}, 247 (1981). 

\end{enumerate}

\newpage
\section*{figure captions}

\begin{figure}
\begin{picture}(300,300)
\put(50,170){\includegraphics[width=15cm]{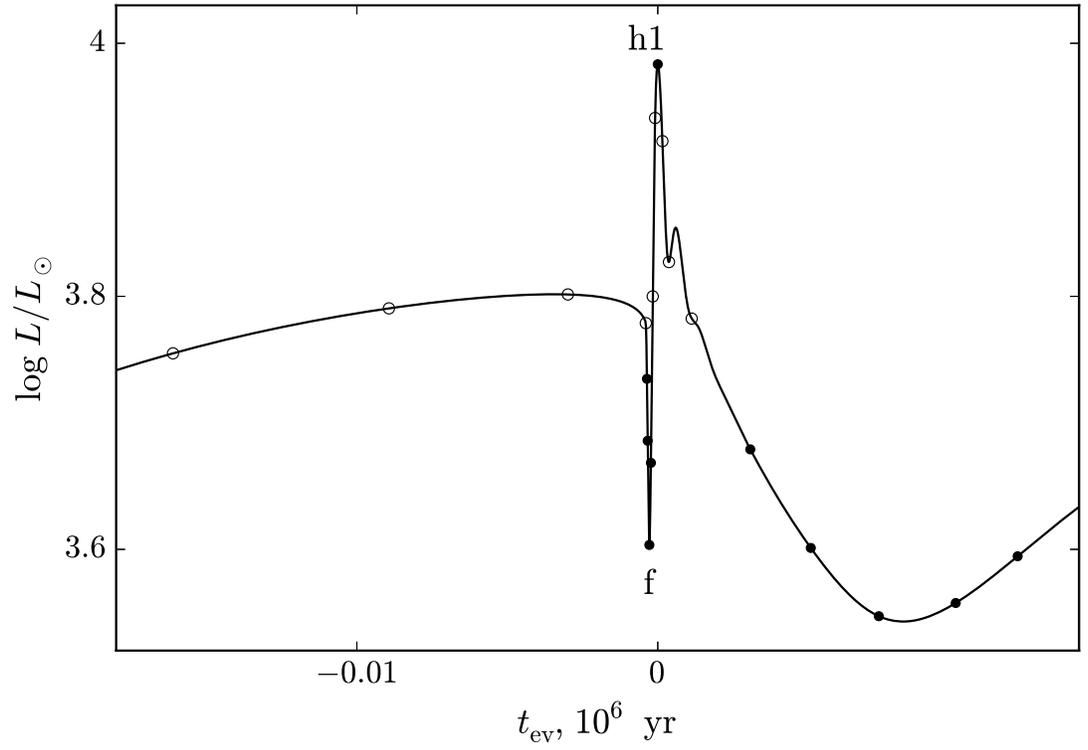}}
\end{picture}
\caption{The stellar luminosity as a function of evolutionary time
in vicinity of the first thermal flash in the helium burning shell.
Pulsationally unstable models ($\eta>0$) are shown in filled cicles
and models with decaying oscillations ($\eta<0$) are shown in open
cicrcles.
The symbols f and h1 indicate the models pulsating in the fundamental
mode and the first overtone, respectively.
}
\label{fig3}
\end{figure}

\begin{figure}
\centerline{\includegraphics[width=15cm]{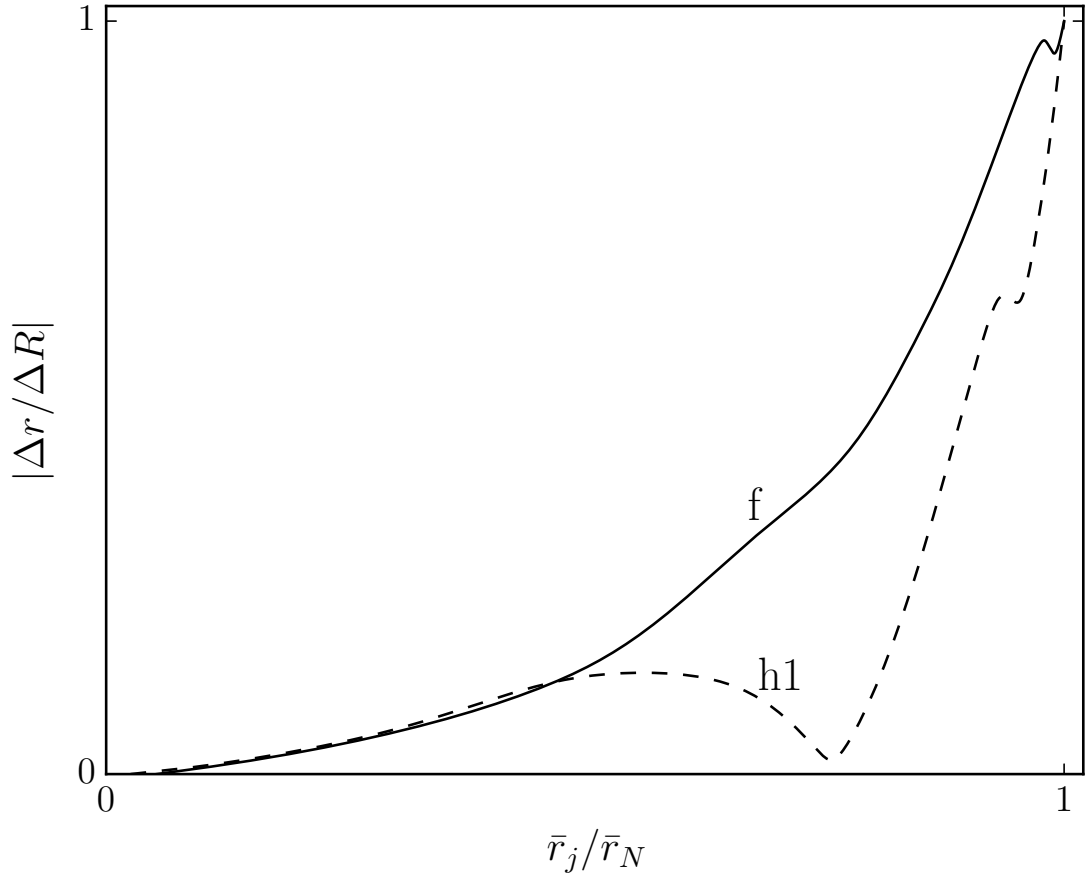}}
\caption{The normalized power spectral density of the radial displacement
at the frequency of the principal pulsation mode as a function of the
mean radial distance.
The plotted radial dependences are calculated for models indicated in
Fig.~\ref{fig3} as f (pulsations in the fundamental mode)
and h1 (pulsations in the first overtone).}
\label{fig4}
\end{figure}

\begin{figure}
\begin{picture}(300,300)
\put(50,170){\includegraphics[width=15cm]{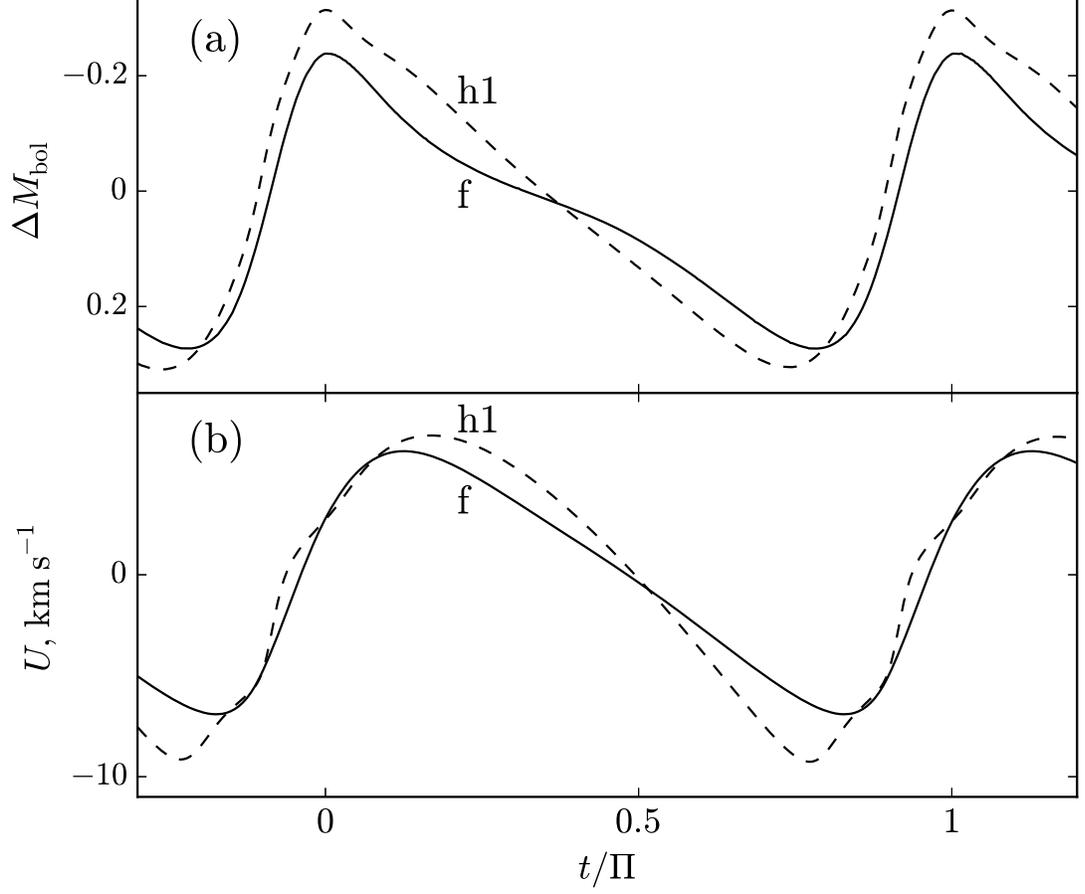}}
\end{picture}
\caption{Temporal dependeces of the bolometric light $\Delta M_\mathrm{bol}$ (a)
and the gas flow velocity at the outer boundary $U$ (b)
for models of stars pulsating in the fundamental mode (f)
and the first overtone (h1).}
\label{fig5}
\end{figure}

\begin{figure}
\begin{picture}(300,300)
\put(50,170){\includegraphics[width=15cm]{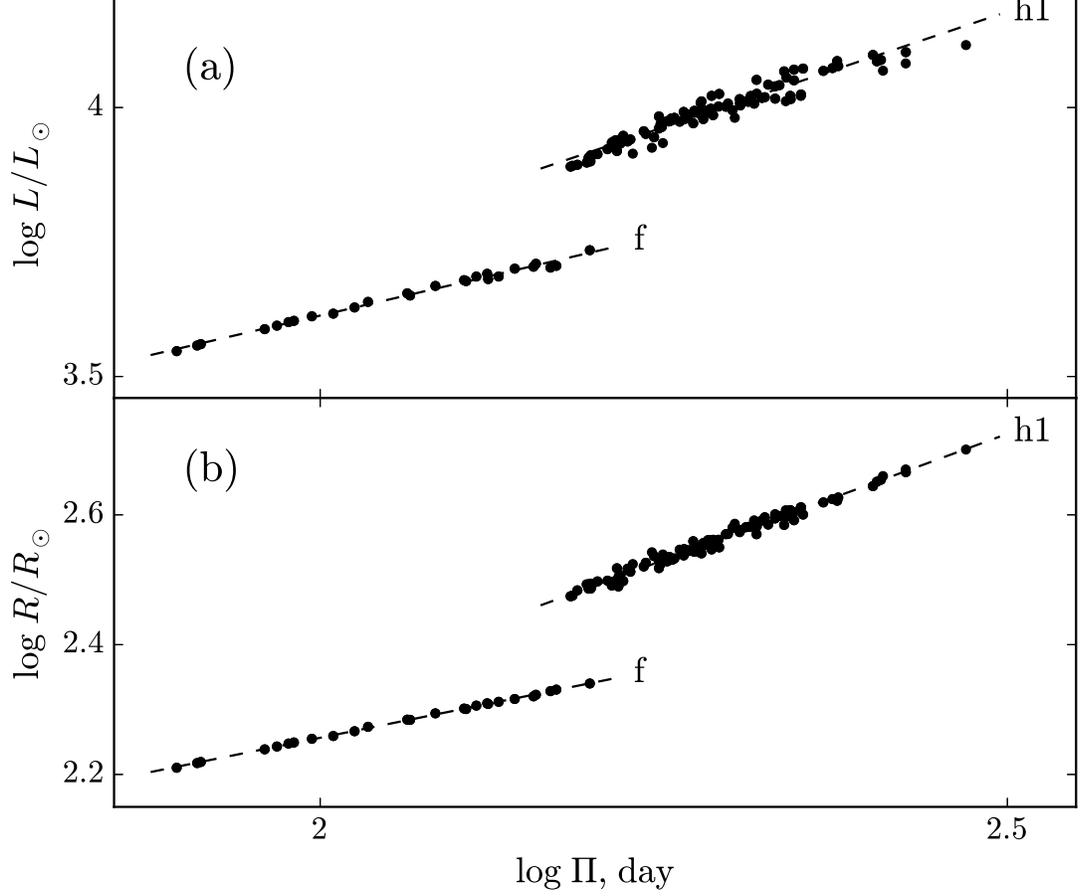}}
\end{picture}
\caption{The period--luminosity (a) and the period--radius (b)
relations for hydrodynamical models of stars pulsating in the fundamental mode
(f) and the first overtone (h1).
Linear fits (\ref{p-l_f}) -- (\ref{p-r_h1}) are shown in dashed lines.}
\label{fig6}
\end{figure}

\begin{figure}
\begin{picture}(300,300)
\put(50,170){\includegraphics[width=15cm]{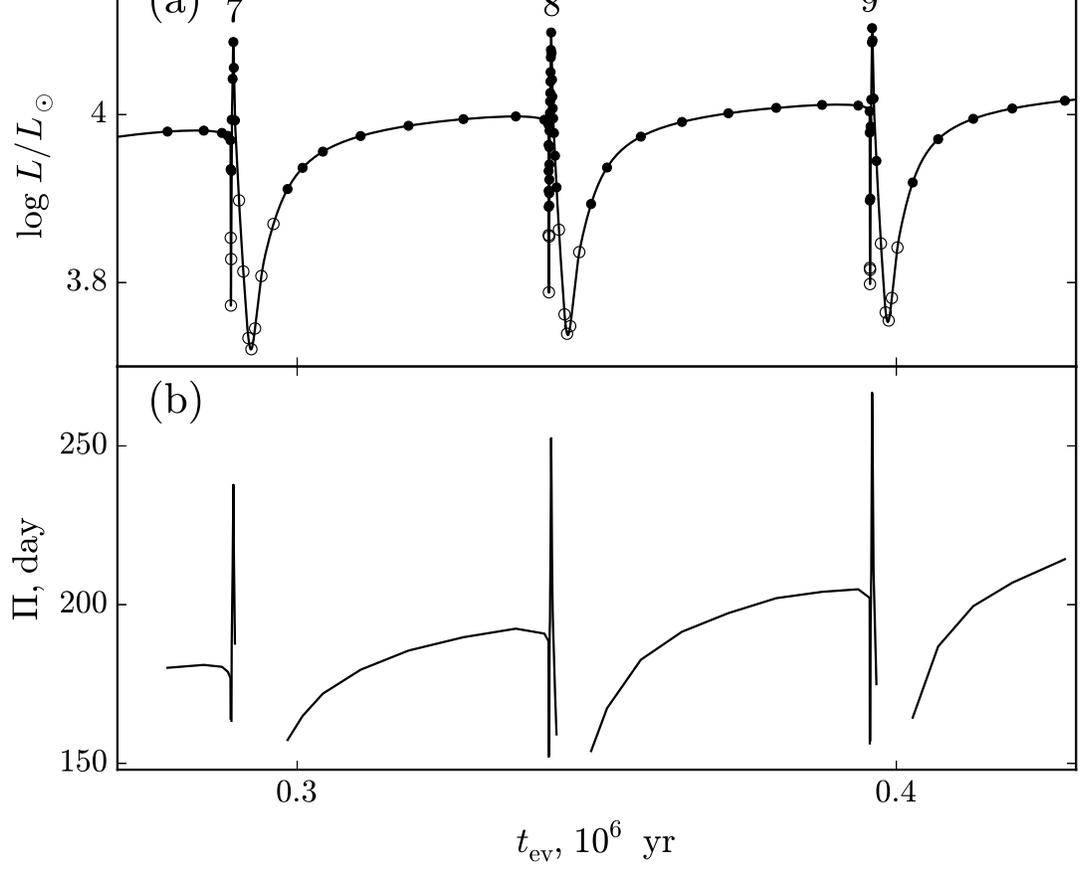}}
\end{picture}
\caption{The stellar luminosity $L$ (a) and the pulsation period $\Pi$ (b)
as a function of the evolutionary time $\tev$.
The number of the thermal flash is indicated at the luminosity maximum.
The filled and open circles show the mean luminosities of
hydrodynamical models that are unstable ($\eta>0$) and stable ($\eta<0$)
against radial oscillations.}
\label{fig7}
\end{figure}

\begin{figure}
\begin{picture}(300,300)
\put(50,170){\includegraphics[width=15cm]{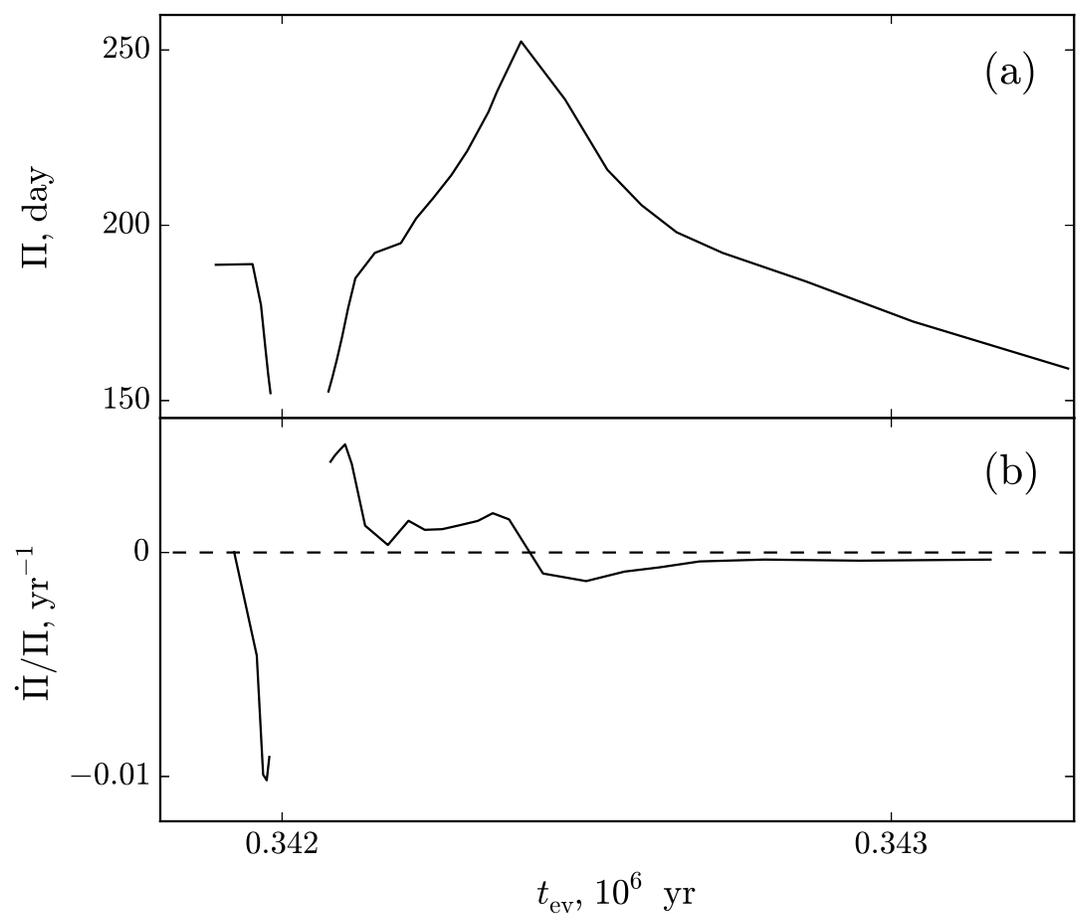}}
\end{picture}
\caption{Temporal dependences of the pulsation period $\Pi$ (a)
and the period change rate $\dot\Pi/\Pi$ (b)
in vicinity of the luminosity maximum of the eight thermal flash.}
\label{fig8}
\end{figure}

\begin{itemize}

\item[Fig. 1.]
The evolutionary track of the star $\mzams=3M_\odot$ from the
         zero age main sequence to the thermally pulsing AGB.

\item[Fig. 2.]
Temporal dependences of the stellar luminosity (a) and the ratio
         of surface mass abundances of carbon and oxygen (b) during
         the AGB stage.

\item[Fig. 3.]
The stellar luminosity as a function of evolutionary time
in vicinity of the first thermal flash in the helium burning shell.
Pulsationally unstable models ($\eta>0$) are shown in filled cicles
and models with decaying oscillations ($\eta<0$) are shown in open
cicrcles.
The symbols f and h1 indicate the models pulsating in the fundamental
mode and the first overtone, respectively.

\item[Fig. 4.]
The normalized power spectral density of the radial displacement
at the frequency of the principal pulsation mode as a function of the
mean radial distance.
The plotted radial dependences are calculated for models indicated in
Fig.~\ref{fig3} as f (pulsations in the fundamental mode)
and h1 (pulsations in the first overtone).

\item[Fig. 5.]
Temporal dependeces of the bolometric light $\Delta M_\mathrm{bol}$ (a)
and the gas flow velocity at the outer boundary $U$ (b)
for models of stars pulsating in the fundamental mode (f)
and the first overtone (h1).

\item[Fig. 6.]
The period--luminosity (a) and the period--radius (b)
relations for hydrodynamical models of stars pulsating in the fundamental mode
(f) and the first overtone (h1).
Linear fits (\ref{p-l_f}) -- (\ref{p-r_h1}) are shown in dashed lines.

\item[Fig. 7.]
The stellar luminosity $L$ (a) and the pulsation period $\Pi$ (b)
as a function of the evolutionary time $\tev$.
The number of the thermal flash is indicated at the luminosity maximum.
The filled and open circles show the mean luminosities of
hydrodynamical models that are unstable ($\eta>0$) and stable ($\eta<0$)
against radial oscillations.

\item[Fig. 8.]
Temporal dependences of the pulsation period $\Pi$ (a)
and the period change rate $\dot\Pi/\Pi$ (b)
in vicinity of the luminosity maximum of the eight thermal flash.

\end{itemize}

\newpage
\clearpage

\newpage
\clearpage

\newpage
\clearpage

\newpage
\clearpage

\newpage
\clearpage

\newpage
\clearpage

\newpage
\clearpage

\newpage
\clearpage

\end{document}